\documentclass{aastex61}

\usepackage{CJK}
\usepackage{amssymb}
\usepackage{multirow}
\usepackage{morefloats}
\usepackage{amsmath}
\usepackage{bigstrut}
\usepackage{booktabs}
\usepackage{natbib}
\usepackage{float}
\usepackage{graphicx,epsfig,fancyhdr,epsf,txfonts,epstopdf}
\usepackage{latexsym,bbm}
\usepackage{lineno}
\usepackage{color}
\usepackage{ulem}
\usepackage{threeparttable}
\usepackage{longtable}
\usepackage{hyperref}

\usepackage{lineno}

\received{***}
\revised{***}
\accepted{***}

\shortauthors{Li et al.}
\shorttitle{Simulation of DPR and Zebras}
\begin{document}
\large
\title{PIC Simulation of Double Plasma Resonance and Zebra Pattern of Solar Radio Bursts}

\correspondingauthor{Yao Chen}
\email{yaochen@sdu.edu.cn}

\author{Chuanyang Li}
\affiliation{Center for Integrated Research on Space Science, Astronomy, and Physics, Institute of Frontier and Interdisciplinary Science, Shandong University, Qingdao, Shandong, 266237, China}
\affiliation{Institute of Space Sciences, Shandong University, Weihai, Shandong 264209, China}

\author{Yao Chen}
\affiliation{Institute of Space Sciences, Shandong University, Weihai, Shandong 264209, China}
\affiliation{Center for Integrated Research on Space Science, Astronomy, and Physics, Institute of Frontier and Interdisciplinary Science, Shandong University, Qingdao, Shandong, 266237, China}

\author{Sulan Ni}
\affiliation{Institute of Space Sciences, Shandong University, Weihai, Shandong 264209, China}
\affiliation{Center for Integrated Research on Space Science, Astronomy, and Physics, Institute of Frontier and Interdisciplinary Science, Shandong University, Qingdao, Shandong, 266237, China}

\author{Baolin Tan}
\affiliation{Key Laboratory of Solar Activity, National Astronomical Observatories of Chinese Academy of Sciences, Beijing, China}

\author{Hao Ning}
\affiliation{Institute of Space Sciences, Shandong University, Weihai, Shandong 264209, China}
\affiliation{Center for Integrated Research on Space Science, Astronomy, and Physics, Institute of Frontier and Interdisciplinary Science, Shandong University, Qingdao, Shandong, 266237, China}

\author{Zilong Zhang}
\affiliation{Institute of Space Sciences, Shandong University, Weihai, Shandong 264209, China}
\affiliation{Center for Integrated Research on Space Science, Astronomy, and Physics, Institute of Frontier and Interdisciplinary Science, Shandong University, Qingdao, Shandong, 266237, China}


\begin{abstract}
\large
Latest study reports that plasma emission can be generated by energetic electrons of DGH distribution via the electron cyclotron maser instability (ECMI) in plasmas characterized by a large ratio of plasma oscillation frequency to electron gyro-frequency ($\omega_{pe}/\Omega_{ce}$). In this study, on the basis of the ECMI-plasma emission mechanism, we examine the double plasma resonance (DPR) effect and the corresponding plasma emission at both harmonic (H) and fundamental (F) bands using PIC simulations with various $\omega_{pe}/\Omega_{ce}$. This allows us to directly simulate the feature of zebra pattern (ZP) observed in solar radio bursts for the first time. We find that (1) the simulations reproduce the DPR effect nicely for the upper hybrid (UH) and Z modes, as seen from their variation of intensity and linear growth rate with $\omega_{pe}/\Omega_{ce}$, (2) the intensity of the H emission is stronger than that of the F emission by $\sim$ 2 orders of magnitude and vary periodically with increasing $\omega_{pe}/\Omega_{ce}$, while the F emission is too weak to be significant, therefore we suggest that it is the H emission accounting for solar ZPs, (3) the peak-valley contrast of the total intensity of H is $\sim 4$, and the peak lies around integer values of $\omega_{pe}/\Omega_{ce}$ (= 10 and 11) for the present parameter setup. We also evaluate the effect of energy of energetic electrons on the characteristics of ECMI-excited waves and plasma radiation. The study provides novel insight on the physical origin of ZPs of solar radio bursts.
\end{abstract}

\keywords{
\href{http://astrothesaurus.org/uat/1261}{Plasma astrophysics (1261)};
\href{http://astrothesaurus.org/uat/1339}{Radio bursts (1339)};
\href{http://astrothesaurus.org/uat/1483}{Solar corona (1483)};
\href{http://astrothesaurus.org/uat/1475}{Solar activity (1475)};
\href{http://astrothesaurus.org/uat/1993}{Solar coronal radio emission (1993)};
}

\section{Introduction}
Zebra Patterns (ZPs) represent a kind of spectral fine structure with equidistant or almost-equidistant stripes of enhanced intensity against a broadband background, frequently observed in dynamic spectra of solar radio bursts such as type IVs (\citealp{Slottje1972, Chernov12}). ZPs receive lots of attention not only because their intriguing and perplexing manifestation, but also because their scientific values in diagnosing coronal parameters such as magnetic field strength within the source of radio bursts. Many models of ZPs have been proposed (\citealp{Kuijpers1975a, Kuijpers1975b, Zheleznyakov1975, Chernov1990, LaBelle2003, Kuznetsov2005, Ledenev2006, Tan2010, Karlicky2013}; see \citealp{Chernov2011} and \citealp{Tan14} for a review). Among them, the well-accepted one is the so-called double plasma resonance (DPR) model. The effect of DPR is associated with plasma kinetic instability excited by energetic electrons in the parameter regime of $\omega_{pe}/\Omega_{ce}>>1$ where $\omega_{pe}/\Omega_{ce}$ is the ratio of plasma frequency $\omega_{pe}$ to electron gyro-frequency $\Omega_{ce}$. It results in sharply-increased growth rates of plasma waves such as the upper hybrid (UH) mode when the UH frequency ($\omega_{UH}= \sqrt{\omega^2_{pe} + \Omega^2_{ce}}$) is equal to $s$ times of $\Omega_{ce}$ ($ \omega_{UH} \approx s \Omega_{ce}$), where $s$ is an integer (\citealp{Zheleznyakov1975, Zlotnik13}, see \citealp{Zheleznyakov2016} for a latest review on DPR).

Previous studies on DPR are mostly linear or quasi-linear analysis of growth rates of electrostatic UH modes (see two latest studies by \citealp{Benacek17} and \citealp{Li2019}), which are however non-escaping and cannot directly account for bursts of the radio emission. Most studies simply presume that the UH modes can somehow convert to the escaping radiation through the nonlinear mode-coupling process in terms of plasma emission (\citealp{Winglee86, Yasnov04, Benacek17}). To understand the underlying radiation process, it is necessary to employ fully kinetic electromagnetic particle-in-cell (PIC) simulation. Yet, only few such studies exist, and are still limited to investigation on the electrostatic modes. One latest example is the work done by \cite{Benacek2019}, who studied the effect of $ \omega_{pe} / \Omega_{ce} $ on the growth rate of electrostatic UH waves.

It is generally believed that solar radio continuum bursts like type-IVs are associated with energetic electrons trapped within magnetic structures in the corona (or in the solar wind for interplanetary type-IVs, \citealp{Smerd1971,Wild1972,Vlahos1982,Stewart1985,Benz2002}; see \citealp{Vasanth2016, Vasanth2019} for lastest observational studies). Most, if not all, earlier theoretical studies on type-IVs employ loss-cone type distributions, such as the classical loss-cone distribution or the Dory-Guest-Harris distribution (DGH: \citealp{Dory65, Winglee86}, \citealp{Benacek17},\citealp{Benacek18}, \citealp{Yasnov04}, \citealp{Li2019}). Using the DGH distribution, \cite{Ni2020} investigated the plasma emission process driven by energetic electrons via the electron cyclotron maser instability (ECMI) within a parameter regime of $ \omega_{pe} / \Omega_{ce} \gg 1 $, while most PIC studies on the ECMI are conducted in the opposite parameter regime ($ \omega_{pe} / \Omega_{ce} \ll 1 $) within which the well-known electron-cyclotron-maser emission applies (\citealp{Wu1979}). The ECMI-plasma emission process starts from linear excitation of waves (like the UH, Z, and whistler (W) modes) by hot DGH electrons within an appropriate coronal background with $\omega_{pe} / \Omega_{ce} >> 1$, the plasma radiation is suggested to be a result of nonlinear wave-wave coupling or coalescence with the F emission generated through the coupling of almost-counter propagating Z and W modes, and H through the coupling of almost counter-propagating electrostatic UH modes. Resonance conditions can be satisfied as demonstrated by \cite{Ni2020}. This radiation process still belongs to the general classification of plasma emission, but is distinct from the traditional beam-driven process (\citealp{Ginzburg1958}).

This study has two major purposes. One is to further verify the ECMI-plasma emission process proposed by \cite{Ni2020} through detailed parameter study. \cite{Ni2020} only considers one specific parameter with $\omega_{pe}/\Omega_{ce}=10$, here we present PIC simulations with $\omega_{pe}/\Omega_{ce}$ varying within two representative gyro-harmonic bands ($ 9.5-11.5 $) so as to simulate the DPR effect. The DPR effect will yield high and low values of growth rates of modes that are excited linearly, thus relevant radiations shall carry the imprint of these variations if the proposed radiation mechanism indeed works. The other purpose is to simulate the development of the DPR effect from linear excitation of non-escaping wave modes to the release of the escaping fundamental (F) and harmonic (H) plasma radiations. This is one necessary step towards a better understanding of ZPs of solar radio bursts such as type-IVs, and done for the first time to our knowledge. The following section introduces the PIC code and the parameter setup. In Section 3 results of the parameter study are presented, followed by the summary and discussion.

\section{The PIC Code and Parameter Setup of Simulations}
The numerical simulation is performed using the Vector-PIC (VPIC) code developed and released by Los Alamos National Labs. VPIC employs a second-order, explicit, leapfrog algorithm to update charged particle positions and velocities in order to solve the relativistic kinetic equation for each species, along with a full Maxwell description for electric and magnetic fields evolved via a second-order finite-difference time-domain solver (\citealp{Bowers2008a,Bowers2008b, Bowers2009}).

The background magnetic field is set to be $ \vec{B}_0 \ (= B_0 \hat{e}_z) $, and the wave vector $ \vec{k} $ is in the $ xOz $ plane, so $ E_y $ represents the pure transverse component of the wave electric field. Periodic boundary conditions are used. The plasmas consist of three components, including background electrons and protons with an Maxwellian distribution, and energetic electrons with the DGH distribution ($ j = 1 $) expressed as follows:
\begin{equation}
f_0=\frac{1}{(2\pi)^{3/2}v_0^{3}}\exp\left(-\frac{u^{2}}{2v_0^{2}}\right), f_e=\frac{u_\perp^{2j}}{2^j(2\pi)^{3/2}v_e^{3+2j}j!}\exp\left(-\frac{u_\perp^{2}+u_\parallel^{2}}{2v_e^{2}}\right)
\label{E4.1}
\end{equation}
where $ u_x $, $ u_y $, $ u_z $ are momentum per mass of particles, $v_0=\sqrt{k_BT_0/m_e}$ is the thermal velocity of background electrons with a fixed value of $ v_0 = 0.018 c \ (T_0\sim 2 \ \rm MK) $, and $v_e$ is the mean velocity of energetic electrons.
The initial plasma temperature of protons is set to be equal to that of background electrons.
Value of $ v_e $ will be adjusted as presented in the following section. All particles distribute homogeneously in space.

The simulation domain is set to be $L_x=L_z=1024 \ dx=1024 \ dz$, where $dx=dz=3.25 \ \lambda_D$ is the grid spacing (or cell size), and $\lambda_D$ is the Debye length of the background electrons. The unit of time is the plasma response time ($\omega^{-1}_{pe}$). The wave number range that can be resolved is [-536, 536] $\Omega_{ce}/c$, and the resolvable frequency range is [0, 32] $\Omega_{ce}$. The resolution in wave number is $\sim1.04 \ \Omega_{ce}/c$, and the resolution in frequency is $\sim0.06 \ \Omega_{ce}$ (for the time interval of 1000 $\omega_{pe}^{-1}$). The simulation time is 3500 $ \omega ^ {-1} _ {pe} $. The time step $dt$ is $0.7 \ dx /(\sqrt{2}c) \sim 0.03 \ \omega_{pe}^{-1}$, in accordance with the Courant condition. The NPPC (number of macro-particles for each species in each cell) is taken to be 2000 for the study on the effect of $v_e$ (\S\ref{sec:3.1}), and 1000 for the larger set of parameter study on the effect of $ \omega_{pe} / \Omega_{ce} $ to reduce the computational cost (\S\ref{sec:3.2} and \S\ref{sec:3.3}). Charge neutrality is maintained initially. The proton-to-electron mass ratio of 1836 is used, and the number density ratio of DGH to total electrons is assumed to be 0.1.

\section{Numerical Results}
According to the latest theoretical analysis of ECMI by \cite{Li2019}, the peak-bottom contrast of growth rate for UH or Z mode is in general larger for smaller $v_e$, thus to reveal as-large-as-possible contrast of wave excitation within the chosen harmonic bands, we set $v_e = 0.15 c$. This corresponds to the lowest value investigated by \cite{Li2019}. To compare with the already-published results for $v_e = 0.3 c$ (\citealp{Ni2020}), we start from a parameter study on $v_e$ while keeping other parameters and configurations the same as those used in \cite{Ni2020}. Two values of $v_e$, $0.15c$ and $0.4c$, are employed. This gives us three cases for comparison, allowing us to evaluate the effect of $v_e$ on the wave growth and emission properties.

\subsection{Effect of $v_e$ on Wave Growth and Emission Properties}
\label{sec:3.1}
\begin{figure*}
   \centerline{\includegraphics[width=0.95\textwidth]{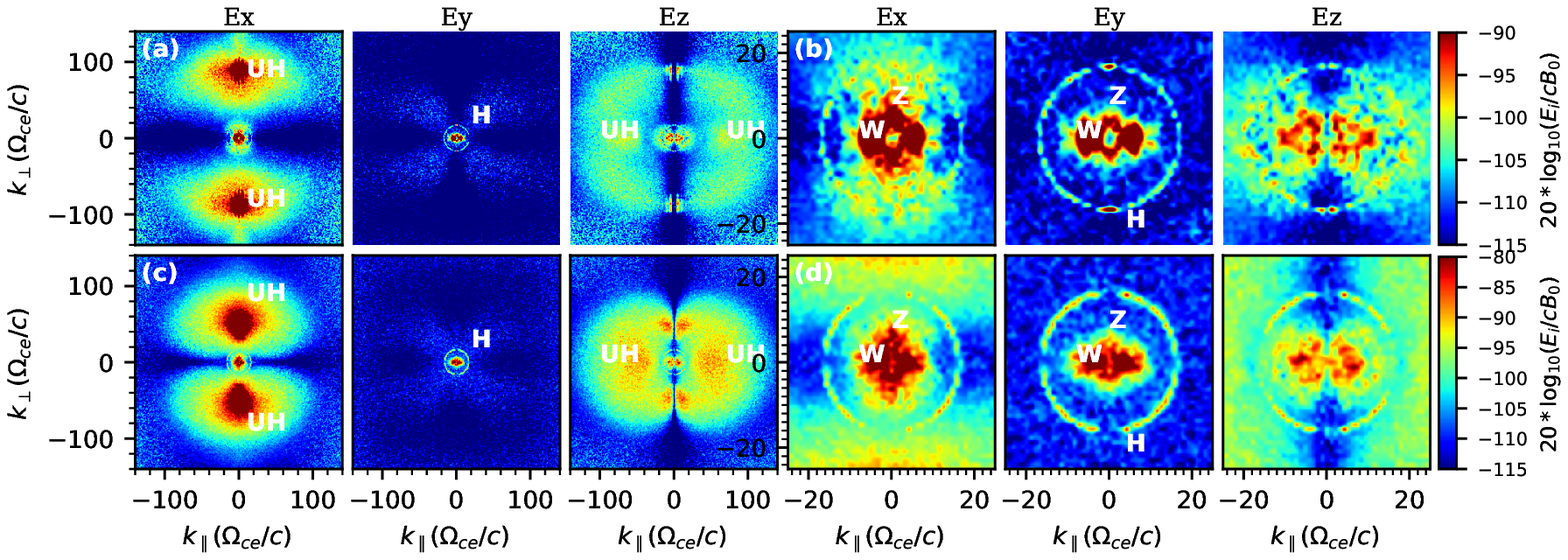}
              }
              \caption{Maximum intensity of ($E_x, E_y, E_z$) in the $\omega$ domain as a function of $k_\parallel$ and $\ k_\perp$ over the interval of $2500<\omega_{pe}t<3500$, as shown by the colormap of $20\log_{10}[(E_x, E_y, E_z)/(cB_0)]$, (a-b) for $v_e=0.15 c$, (c-d) for $v_e=0.4 c$, panels (b) and (d) are zoom-in versions of panels (a) and (c). ``UH'' stands for upper hybrid mode, ``W'' for whistler mode, ``Z'' for Z mode, and ``H'' for harmonic plasma emission.
              }
   \label{Fig:figure1}
   \end{figure*}

\begin{figure*}
   \centerline{\includegraphics[width=0.95\textwidth]{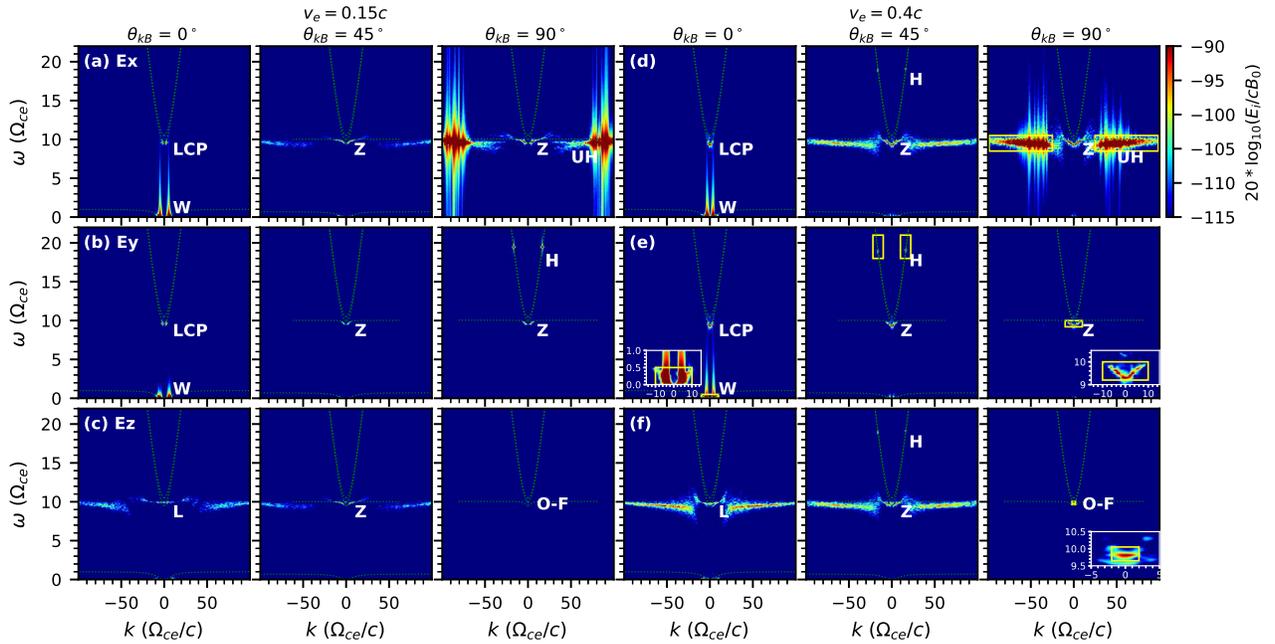}
              }
              \caption{Wave dispersion diagrams of $E_x$, $E_y$, and $E_z$ over times $2500 < \omega_{pe}t < 3500 $ along directions of $0^\circ, 45^\circ$ and $90^\circ $ (the angle between wave vector and background magnetic field). Left panels (a-c) for $v_e=0.15 c$ while right panels (d-f) for $v_e=0.4 c$. ``L'' stands for Langmuir mode, ``LCP'' for left-circularly polarized wave and ``O-F'' for O mode around the fundamental plasma frequency. The green dotted lines represent the dispersion curves of four modes (X, O, Z and W) of the  theoretical magneto-ionic theory. The yellow boxes mark the spectral region to calculate the intensity of different wave modes.
              }
   \label{Fig:figure2}
   \end{figure*}

Figure \ref{Fig:figure1} shows maps of the maximum wave intensity in the $\vec k \ (k_\parallel, k_\perp)$ space during the last stage of simulation ([2500, 3500] $\omega^{-1}_{pe}$), and Figure \ref{Fig:figure2} shows the $\omega-k$ dispersion diagram along three different directions. The wave modes exhibited in both figures can be identified with the dispersion curves of the four magneto-ionic modes (X, O, Z, and W) that are superposed onto Figure \ref{Fig:figure2}. It should be noted that the O-F mode is quasi-electrostatic here since its frequency is very close to the plasma oscillation frequency, thus the magnetic field energy is much weaker than the electric field energy. In addition, the UH mode is electrostatic, while the Z mode contains both electrostatic and electromagnetic parts. Since the transverse electromagnetic component can be represented by $E_y$, in this study we use dispersion diagrams of components of the electric field to represent various wave modes.

The large regime with enhanced intensity at large $k$ in Figure \ref{Fig:figure1} belongs to the UH mode (i.e, obliquely-propagating Langmuir wave), which is the extension of the Z mode towards large $k$. Note that a spectral gap exists between the electrostatic UH and the electromagnetic Z mode (see Figure \ref{Fig:figure2}). For $v_e = 0.15 c$ the gap is in the range of $ 20 \ \Omega_{ce} / c < |k| < 50 \ \Omega_{ce} / c $, and for $v_e=0.4 c$ the gap is within $ 10 \ \Omega_{ce} / c <|k| < 20 \ \Omega_{ce} / c $. The circular pattern shown in Figure \ref{Fig:figure1} belongs to the H emission, inside the circle are W and Z modes with smaller $k$. In addition, the W mode is mainly along quasi-parallel to parallel direction and Z mode mainly along oblique to perpendicular direction. The O-mode F emission is too weak to be recognized from Figure \ref{Fig:figure1}.

\begin{figure*}
   \centerline{\includegraphics[width=0.90\textwidth]{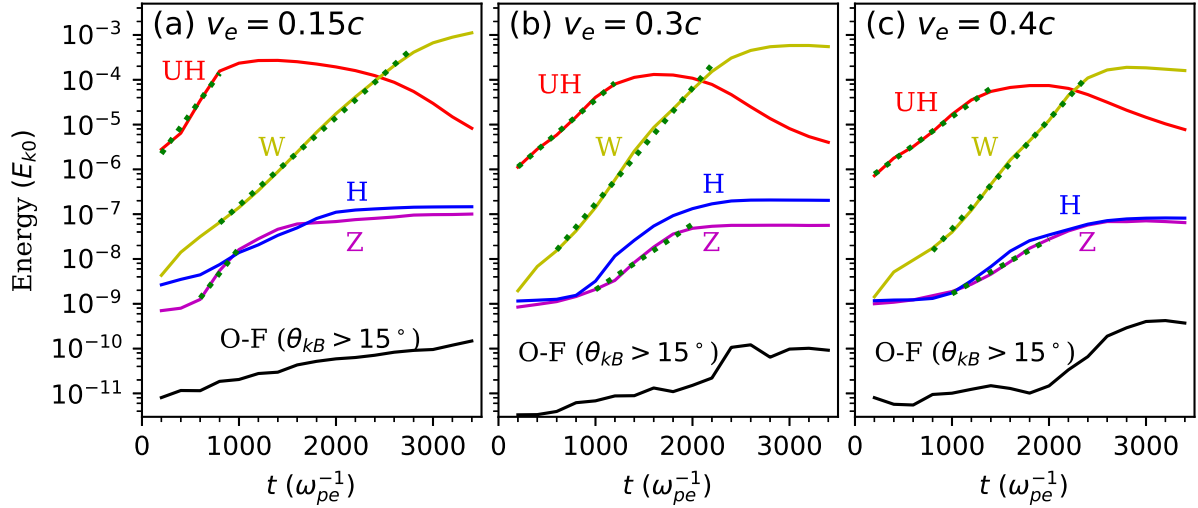}
              }
              \caption{
              The temporal energy profiles of UH, Z, W, H, O-F modes with $\omega_{pe} / \Omega_{ce} =10$ and (a) $v_e=0.15c$, (b) $v_e=0.3c$ and (c) $v_e=0.4c$, normalized to the respective initial kinetic energy of total electrons ($E_{k0}$). Among them, the energy of O-F mode are integrated in the range of $\theta_{kB} > 15^\circ$ to avoid contamination  from nearby
              Langmuir wave. The three dotted lines in each panel represent exponential fittings to energy profiles.
              }
   \label{Fig:figure3}
   \end{figure*}

Figure \ref{Fig:figure3} presents energy profiles of these modes, which are calculated by integrating energy of the six components of the electro-magnetic field within respective range of the dispersion curve as indicated by spectral boxes overplotted in Figure \ref{Fig:figure2}, according to the Parseval's theorem.
The UH, W, and H modes are well separated from each other in either frequency or wave number. Yet, it is difficult to separate the O-F mode from the Z mode along parallel and quasi-parallel propagating directions along which the two modes are somehow connected. We have therefore excluded this part ($\theta_{kB}<15^\circ$) when calculating the energy of the O-F mode. Note that, the thermal noise, even organized along the dispersion curves, presents minor contribution to the total energy of the excited wave modes (see more details in the Appendix).


As seen from Figure \ref{Fig:figure3}, there exist three major consequences as $v_e$ increases from $0.15 c$ to $0.4 c$.

First, change of the angular distribution of the UH mode in the $\vec k$ space. For $v_e = 0.15 c$, the strongest part of UH grows mainly along quasi-perpendicular to perpendicular direction within a quite-limited range of propagation angles ($\sim 82 - 98^\circ$), the range of $k$ is also quite limited ($\sim 60-100 \ \Omega_{ce} / c$), both ranges are much smaller than their counterparts for larger $v_e$. For other modes, such as Z and W, the distribution patterns do not change obviously with increasing $v_e$, except the intensities become stronger and the Z mode extends to smaller $k \ (\sim 0)$. Values of $k$ of UH get smaller for larger $v_e$, so as to meet the condition of the wave-particle resonant coupling ($ \omega - n \Omega_{ce} - k_z v_z = 0 $, where $n$ is the harmonic number).

Second, decrease of growth rates of the three linearly-excited modes (UH, Z, and W) with increasing $v_e$. See Figure \ref{Fig:figure3} for lines fitting the linear stage of the mode growth. For $v_e = 0.15 c$ the growth rates of UH, Z, and W modes are 0.34, 0.32, 0.23 $\Omega_{ce}n_e/n_0$, and for $v_e = 0.3 c$ the rates are 0.22, 0.17, 0.28 $\Omega_{ce}n_e/n_0$ (c.f., \citealp{Ni2020}), respectively. For $v_e = 0.4 c$ the growth rates are similar to those of $v_e = 0.3 c$. This variation of growth rate directly affects the duration of the linear stage and the maximum intensity of the mode. In general, the duration of linear stage of UH mode becomes larger with increasing $v_e$, being $\sim1000 \ \omega_{pe}^{-1}$ for $v_e = 0.15 c$ and $\sim1500 \ \omega_{pe}^{-1}$ for both $v_e = 0.3 c$ and $v_e = 0.4 c$. The duration of the Z-mode linear growth also presents similar increasing trend with increasing $v_e$. For the W mode, the duration of the linear stage decreases as $v_e$ increases from $0.15 c$ to $0.3-0.4 c$.

Last, change of directional pattern of the H emission. The results presented here confirm the basic picture first described by \cite{Ni2020} that escaping radiations can be excited by energetic electrons with the DGH distribution. The obtained emission, in particular, the H emission, exhibits quite different characteristics with increasing $v_e$. For $v_e = 0.15 c$ H radiates mainly along the quasi-perpendicular to perpendicular direction, while for larger $v_e$ H exhibits a very-different quadrupolar-like pattern. This difference is likely due to the change of its mother wave, i.e., the UH mode, whose coalescence results in the generation of the H mode according to \cite{Ni2020}. As mentioned, for $v_e = 0.15 c$ the UH mode has a quite-limited range in both propagating angles ($\theta_{kB}$) and value of $k$, this affects the emission pattern of H through the matching condition of wave number; on the other hand, for large $v_e$ the UH mode distributes wider in the $\vec k$ space, this makes the coalescence of UH mode close to those reported for the isotropic Langmuir waves (see, e.g., \citealp{Ziebell2015}) and results in the quadrupolar-like pattern of radiation.

In addition, the intensity of plasma radiations (H or O-F) are comparable for the three cases, despite the significant change of directional pattern of the H mode. The H mode is significantly stronger than the O-F mode ($\theta_{kB} > 15^\circ$) by about two-three orders of magnitude, while the O-F mode is only marginally stronger than the background noise. The H emission concentrates around frequency of $\sim 19.4 \ \Omega_{ce} $ for $v_e =0.15c$ and $ \sim 19.0 \ \Omega_{ce} $ for $v_e =0.4c$.

\subsection{The DPR Effect}
\label{sec:3.2}
To simulate the DPR effect, we vary $\omega_{pe}/\Omega_{ce}$ within two harmonic bands ($9.5-11.5$) with fixed value of $v_e \ (= 0.15 c)$. This allows us to observe two complete periods of the DPR effect which leads to high and low values of growth rates and wave intensities. The basic underlying assumption is that the spatial-temporal scales of variation of $\omega_{pe}/\Omega_{ce}$ should be much larger than corresponding scales of wave excitation. This allows us to use PIC simulations within homogeneous background (using different $\omega_{pe}/\Omega_{ce}$) to simulate the DPR effect which rises only in a non-uniform media.

\begin{figure*}
   \centerline{\includegraphics[trim=0cm -1cm 0cm -1cm,scale=0.99]{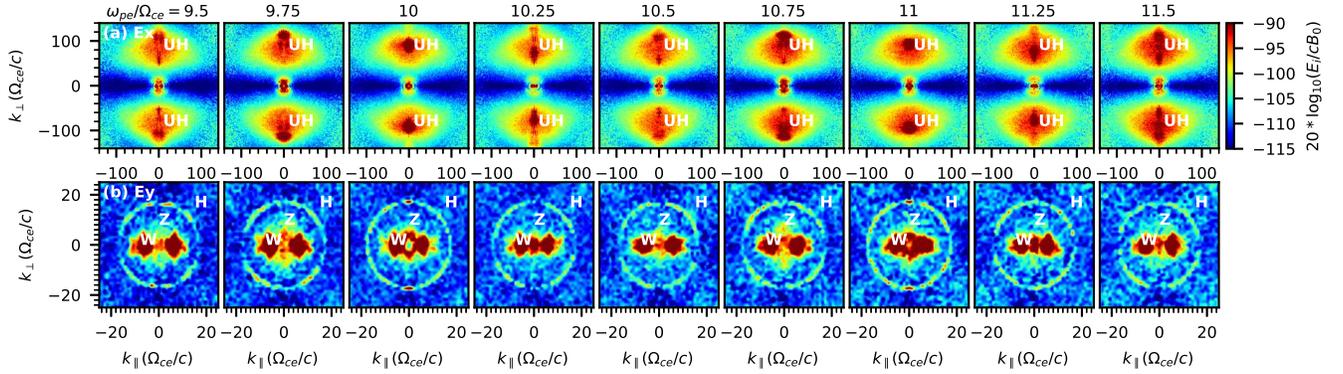} 
              }
              \caption{
              The wave intensity map in ($k_\parallel,k_\perp$) space with $\omega_{pe}/\Omega_{ce} = 9.5 - 11.5$ over the interval of $1500 < \omega_{pe}t < 2500$. Lower panels are zoom-in versions of upper panels.
              }
   \label{Fig:figure4}
   \end{figure*}

\begin{figure*}
   \centerline{\includegraphics[width=0.90\textwidth]{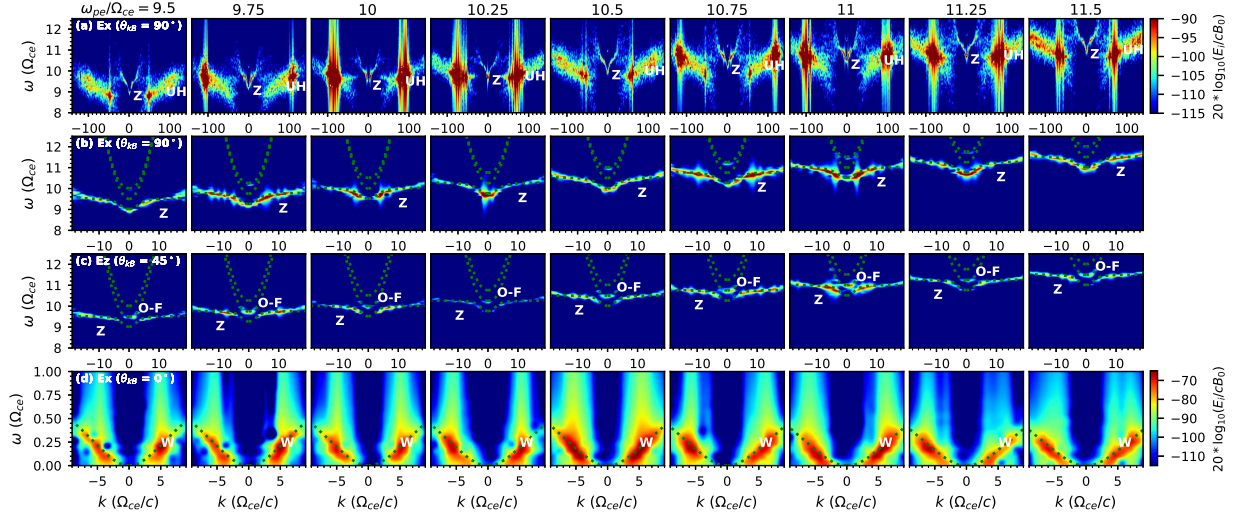}
              }
              \caption{
              The $\omega-k$ dispersion curves for different modes over times $2500 < \omega_{pe}t < 3500$ along the directions labeled on the corresponding panels. Each column represents result for one specific value of $\omega_{pe}/\Omega_{ce}$. The green dotted lines represent the dispersion curves of four modes (X, O, Z and W) of the  theoretical magneto-ionic theory. Panels (a-c) use the same colorbar (next to panel a), and panel (d) use the colorbar next to it.
              }
   \label{Fig:figure5}
   \end{figure*}

Figure \ref{Fig:figure4} presents the wave intensity map in the $\vec k$ space, and Figure \ref{Fig:figure5} presents the $\omega - k$ dispersion analysis. From left to right, results with various $\omega_{pe}/\Omega_{ce}$ in the range of [9.5, 11.5] are presented. As seen from the dispersion analysis shown in Figure \ref{Fig:figure4}a, varying $\omega_{pe}/\Omega_{ce}$ mainly affects the distribution (in $\vec k$ space) of the UH mode with large intensity. Such distribution also oscillates with varying $\omega_{pe}/\Omega_{ce}$ as clearly seen from the figure. For example, when $\omega_{pe}/\Omega_{ce}$ is close or equal to 10 or 11, the range of $k$ is constrained to very-narrow ranges along the perpendicular to quasi-perpendicular direction, consistent with the $\omega-k$ diagram (see Figure \ref{Fig:figure5}a). For the Z mode, when $\omega_{pe}/\Omega_{ce}$ = 10 or 11, no wave exists for small $k$ along the perpendicular direction (Figure \ref{Fig:figure5}b), while for $\omega_{pe}/\Omega_{ce}$ = 10.25 and 11.25 the mode mainly distributes around region of small $k$. For other non-integer values of $\omega_{pe}/\Omega_{ce}$, it distributes over a wider region of $k$. It's frequency increases with increasing $\omega_{pe}/\Omega_{ce}$, as expected. The ranges of $\omega$ and $k$ of the parallel-propagating W mode do not change considerably with $\omega_{pe}/\Omega_{ce}$ according to Figure \ref{Fig:figure5}d, due to its nature of excitation (i.e., the $n=1$ cyclotron resonance).

\begin{figure*}
   \centerline{\includegraphics[width=0.9\textwidth]{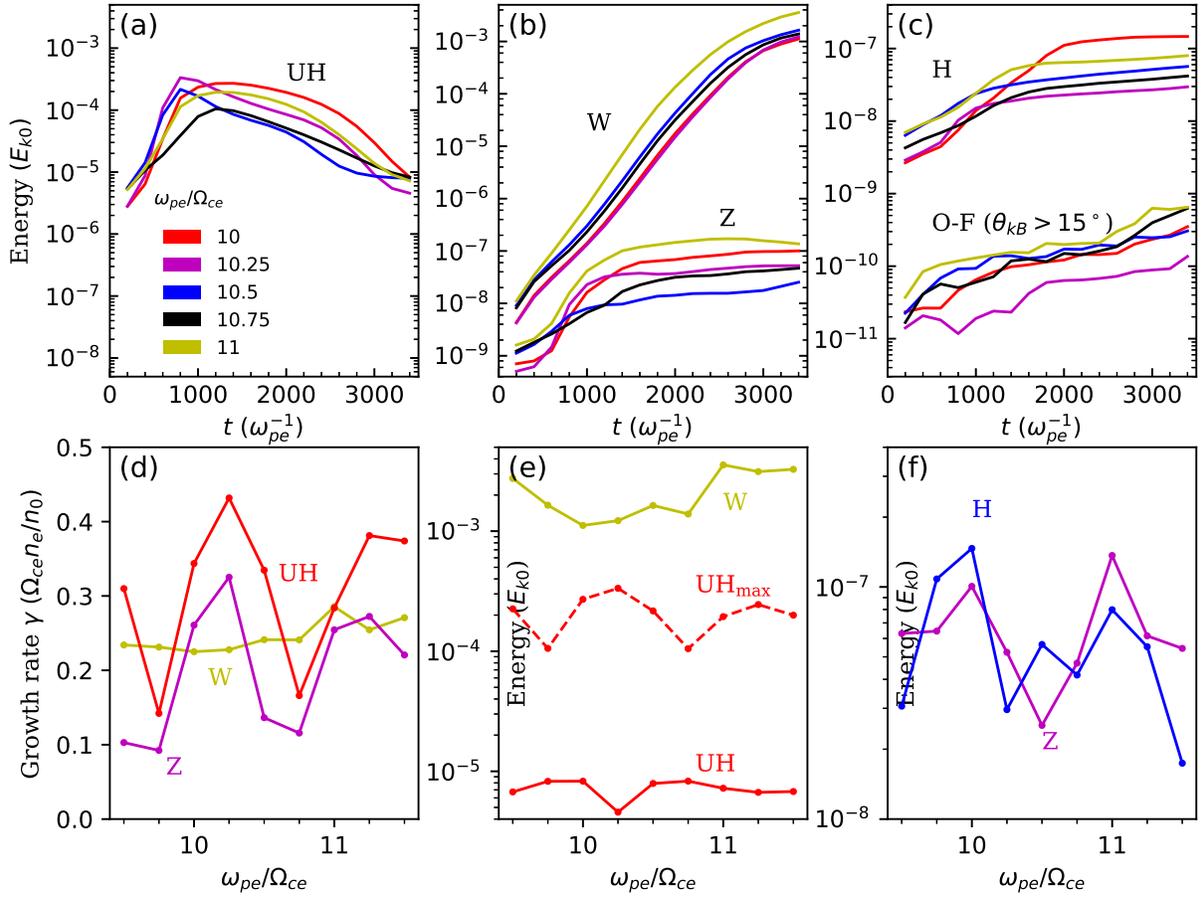}
              }
              \caption{(a-c) The temporal energy profiles of the five modes with $v_e=0.15c$ and $\omega_{pe}/\Omega_{ce} = 10, 10.25, 10.5, 10.75, 11$, normalized to $E_{k0}$. (d) The fitted linear growth rate of UH, Z and W modes. (e-f) The variation of final energy of different modes with $\omega_{pe} / \Omega_{ce}$, the red dashed line represent the maximum energy profile of UH mode.
              }
   \label{Fig:figure6}
   \end{figure*}

The temporal profiles of wave energy have been plotted in the upper panels of Figure \ref{Fig:figure6}, the linear growth rates obtained by exponential fittings have been plotted in panel (d), mode energies at the end of the simulations have been plotted in the last two panels (e-f). In all cases, the UH mode experiences significant damping after its saturation stage, we therefore also plotted the maximum energy of UH mode in Figure \ref{Fig:figure6}e. For other modes, no significant damping is observed.

Consistent with the results presented above, the UH mode dominates the growth of waves during the first 1000 $\omega_{pe}^{-1}$, reaching the maximum energy level around or slightly higher than $10^{-4} \ E_{k0}$ and then damping to an energy level around $10^{-5} \ E_{k0}$. During the saturation and damping stages of UH mode, the W mode becomes the dominant one reaching an energy level around $10^{-3} \ E_{k0}$. The Z mode only reaches a weak level of $10^{-8} - 10^{-7} \ E_{k0}$.

The DPR effect can be clearly seen from the periodic variation of both growth rate and energy of UH and Z modes (see lower panels of Figure \ref{Fig:figure6}). The growth rates manifest variations of two complete periods, reaching the minima at $\omega_{pe}/\Omega_{ce} = $ 9.75 and 10.75 and the maxima at $\omega_{pe}/\Omega_{ce} = $ 10.25 and 11.25. The period of variation is $\Omega_{ce}$ as expected, yet the peaks do not lie exactly at the integer times of $\Omega_{ce}$, this is due to the thermal effect and the deviation from perpendicular propagation of the two modes, as already found earlier by theoretical analysis (see, e.g, \citealp{Winglee86,Yasnov04,Benacek17,Li2019}). The ratio of the maximum to minimum growth rates achieved respectively at $\omega_{pe}/\Omega_{ce} = $ 10.25 and 10.75 is $\sim$ 2.6 for UH mode and $\sim$ 2.8 for Z mode. The growth rate of W mode does not present similar double-period DPR pattern, which is mainly associated with the fundamental gyro-resonance ($n=1$), while the DPR effect is mainly carried by modes (UH and Z) that are excited at large number of harmonics.

The maximum energy of UH mode (indicated as $\rm UH_{max}$ in Figure \ref{Fig:figure6}e) also presents a similar double-period variation, a direct imprint of the DPR effect. Similar trend is observed from the profile of Z mode energy at the end of simulations (Figure \ref{Fig:figure6}f). Yet the energy of UH mode at the end of simulations presents almost-opposite trend of variation, it reaches the minimum level at $\omega_{pe}/\Omega_{ce} = 10.25$ and maximum around $\omega_{pe}/\Omega_{ce} = $ 10.0 and 10.75. The energy of W mode at the end of simulations first decreases and then increases with increasing $\omega_{pe}/\Omega_{ce}$, quite different from the variation of UH and Z modes. This means that variation of $\omega_{pe}/\Omega_{ce}$ does affect the growth of W mode, yet not in accordance with the DPR effect. This is also likely due to the point that the W mode is associated with the $n=1$ gyro-resonance. Note that the W mode has not reached their full saturation level at the end of simulations.

\subsection{The ECMI-induced Plasma Emission with ZPs}
\label{sec:3.3}
According to the PIC simulation and further analysis presented in \cite{Ni2020}, plasma emission of both F (O mode) and H bands can be generated through nonlinear coalescence of waves that are excited by energetic electrons of the DGH distribution. The H emission is generated through coalescence of almost-counter-propagating UH modes and the F emission is generated through coalescence of almost-counter-propagating Z and W modes. Thus, it is intriguing to further examine how the DPR effect affects properties of the obtained plasma emission.

From Figure \ref{Fig:figure6}c, the H emission is stronger than the O-F emission by about 2 orders of magnitude. Further inclusion of the energy of quasi-parallel propagating O-F mode does not affect this result. The final energy of H emission varies within a range of $10^{-8} - 10^{-7} \ E_{k0}$, at the same level of the energy of Z mode; while the O-F mode is only slightly stronger than the background noise as mentioned, within a range of $10^{-10} - 10^{-9} \ E_{k0}$. 
For the H emission, the energy exhibits periodic variation, reaching the local maxima at $\omega_{pe}/\Omega_{ce} = $ 10 and 11 and the local minima at $\omega_{pe}/\Omega_{ce} = 9.5, 10.25$, and 11.5, for the parameters considered here.
This gives rise to the well-known ZP feature of solar radio bursts. The obtained peak-valley ratio of the H intensity is about 4, consistent with the observed range of intensity contrast of solar ZPs (see, e.g., \citealp{Chen2011}, \citealp{Chernov12}, and \citealp{Tan2014}). Note that the variation profile of the H intensity is basically opposite to that of the growth rate and the maximum energy of the UH mode, yet in line with its final energy profile (see Figure \ref{Fig:figure6}e).


For the O-F mode, evaluation of energy may suffer from certain error if considering the contamination from nearby Z-mode wave and its weak intensity. Thus, we do not further analyze the energy variation of the O-F mode with $\omega_{pe}/\Omega_{ce}$. In reality the weak signals of the O-F mode can be easily buried by the strong H emission and thus may not be observable.

From the $\vec k$-space dispersion diagram (Figure \ref{Fig:figure4}b), the H emission exhibits certain variation in directional pattern with $\omega_{pe}/\Omega_{ce}$. When $\omega_{pe}/\Omega_{ce} = $10 and 11, the intensity of the H emission peaks along the perpendicular direction. This is similar to what we have presented in Section \ref{sec:3.1}, likely due to the mentioned specific concentration of the UH mode in the $\vec k$ space. For non-integer values, the H emission exhibits somewhat complex patterns of radiation, it may get enhanced at different propagation angles, and may get weakened along the perpendicular direction (e.g., for cases with $\omega_{pe}/\Omega_{ce} = 9.75, 10.25$, and 11.5). The corresponding distribution of UH modes (see Figure \ref{Fig:figure4}a) is far from isotropic, thus the theory of coalescence of Langmuir waves with isotropic distribution which leads to quadrupolar pattern of radiation does not apply here.


\section{Summary and Discussion}
ZPs are intriguing features observed in the dynamic spectra of solar radio bursts. It carries valuable information of plasmas and magnetic field in the corona and has been used for diagnostic purpose. The DPR effect in plasmas with large $\omega_{pe} / \Omega_{ce}$ represents the most-accepted scenario for ZPs. Yet, earlier studies were mostly based on linear theoretical analysis of electrostatic waves such as the UH mode. This study presents the first PIC study simulating DPR and ZPs simultaneously on the basis of the ECMI-driven plasma emission process.

According to our simulations, both electrostatic UH wave and electromagnetic Z and W modes are excited through ECMI driven by energetic electrons of DGH distribution. Properties of both UH and Z modes are well consistent with the DPR effect. The induced plasma emissions at H and F bands are obtained through nonlinear wave-wave interaction, consistent with our earlier study (\citealp{Ni2020}). The H emission dominates over the F emission in intensity by about 2 orders in magnitude, this indicates that the ZPs arise from the H emission, rather than the F emission as assumed in many studies (e.g., \citealp{Zlotnik2014,Chernov2015,Kaneda2017}). The peak-valley contrast of intensity of the H emission is found to be $\sim$ 4, consistent with some observational reports. These results are critical to understanding of solar ZPs and further diagnostic efforts using radio data.

In addition, we found that the energy (or velocity) of energetic electrons (represented with $v_e$) is important to the characteristics of the UH mode and the radiation pattern of the H emission. For smaller $v_e \ (=0.15 c)$, the UH mode concentrates within a quite-limited range in the $\vec k$ space (mainly along the perpendicular-to-quasi-perpendicular direction), for larger $v_e \ (=0.3 c$ and $0.4 c)$ the UH mode distributes more diffusively over a wider range; the H emission for $v_e \ (=0.15 c)$ achieves the maximum intensity along the perpendicular direction while for larger $v_e$ it presents a quadrupolar-like radiation pattern lack of perpendicular propagation. This result is obtained for fixed value of $\omega_{pe} / \Omega_{ce} \ (= 10)$. Note that varying $\omega_{pe} / \Omega_{ce}$ also affects the distribution and radiation pattern of various wave modes.

In the present study, the DPR effect and the induced plasma emission of ZPs are investigated with various $\omega_{pe} / \Omega_{ce}$ and fixed $v_e \ (= 0.15 c)$. In future, PIC simulations with larger $v_e$ shall be conducted to further explore its effect. In addition, the DPR effect is simulated here with independent PIC simulations for different values of $\omega_{pe} / \Omega_{ce}$, assuming the spatial-temporal scales of inhomogeneity are much larger than the scales of wave excitation. Studies within a larger domain incorporating inhomogeneous distribution of magnetic field strength and plasma density should be carried out so as to simulate the DPR effect and generation of ZPs self-consistently.

\section*{Acknowledgements}
This study is supported by the National Natural Science Foundation of China (11790303 (11790300), 11750110424 and 11873036). The authors acknowledge the Beijing Super Cloud Computing Center (BSCC, URL: http://www.blsc.cn/) for providing high-performance computing (HPC) resources, and the open-source Vector-PIC (VPIC) code provided by Los Alamos National Labs (LANL). The authors acknowledge the helpful discussions with Dr. Xiaocan Li (LANL) and Prof. Quanming Lu (USTC, University of Science and Technology of China). The authors are grateful to the anonymous referee for valuable comments.


\appendix

\section{Convergence test of the PIC simulation}

The convergence test of PIC simulations is computationally expensive. Here we limit the test to verify the significance of escaping radiations and intensity variations of various wave modes with time. Setups of these test cases are summarized in Table \ref{Tab:Table1}.

Note that the number of grid points, NPPC (number of particles per cell), computational domains, and durations of the present study are either much higher than those earlier studies or comparable to those latest studies. For example, \cite{Kasaba2001} used 16 particles per cell for background particles and 4 for the electron beam, while in \cite{Umeda2010} the NPPC is set to be 256, in \cite{Thurgood2015} the NPPC is set to be 1000, and it is set to be 2000 in \cite{Zhou2020}.

Similar convergence test has been done earlier by \cite{Ni2020}, here we present a more-complete test in accordance with the parameters adopted here. These cases are briefly explained below, in comparison to the reference case, which is taken to be the case presented in \S\ref{sec:3.2} with $\omega_{pe}/\Omega_{ce}=10$, $v_e=0.15c$, and NPPC=1000.

\begin{itemize}
  \item Case A: for all particles with the Maxwellian distributions, no energetic particles are involved, to check the level of the organized thermal noise.
  \item Case B: with a larger number of spatial grid points and thus a larger domain, to check the effect of grid number and size of the domain.
  \item Case C: with less NPPC yet the same total number of particles in the same spatial domain, with smaller cell size yet more grid points, to check the effect of NPPC and size of cells.
  \item Cases D-G: with various NPPC from 200 to 4000, to check the effect of NPPC.

\end{itemize}

\begin{figure*}
   \centerline{\includegraphics[width=0.9\textwidth]{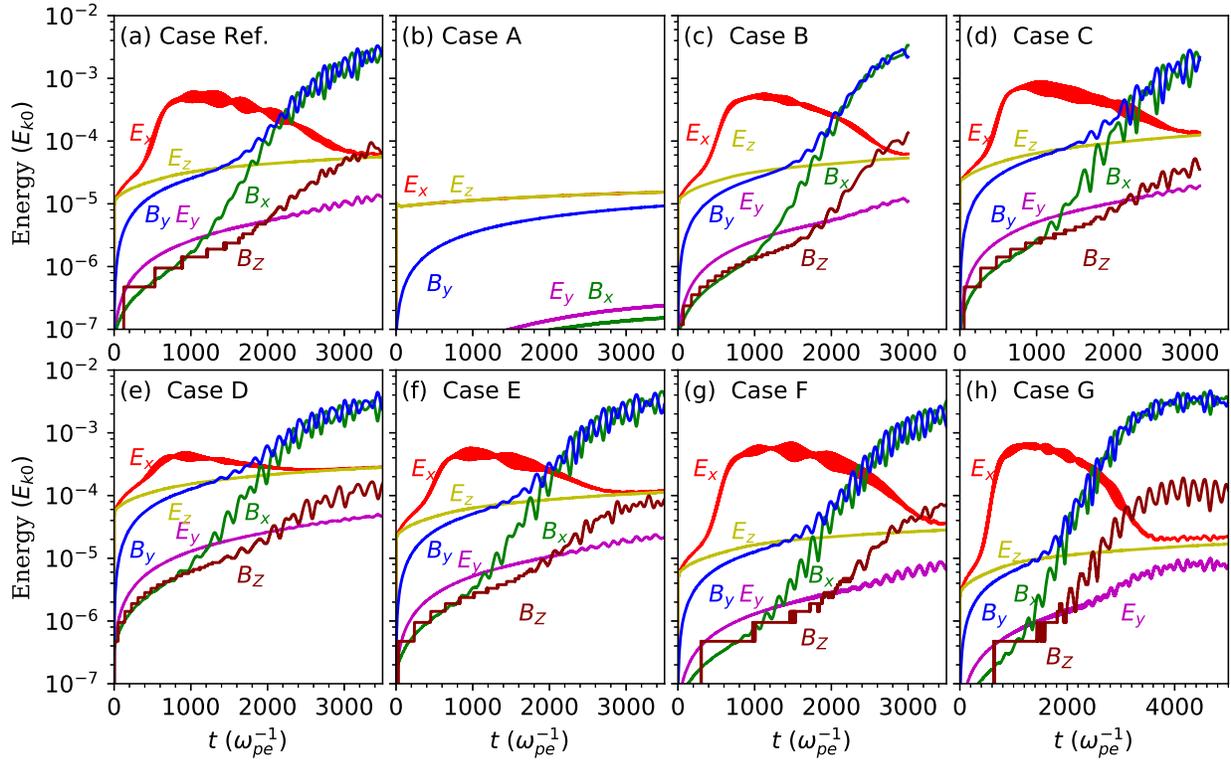}
              }
              \caption{Temporal profiles of energies of various field components ($E_x, E_y, E_z, B_x, B_y$, and $B_z$) normalized to the initial kinetic energy of the DGH electrons ($E_{k0}$). See Table \ref{Tab:Table1} for the setup parameters of each case.
              }
   \label{Fig:figure7}
   \end{figure*}

\begin{figure*}
   \centerline{\includegraphics[width=0.9\textwidth]{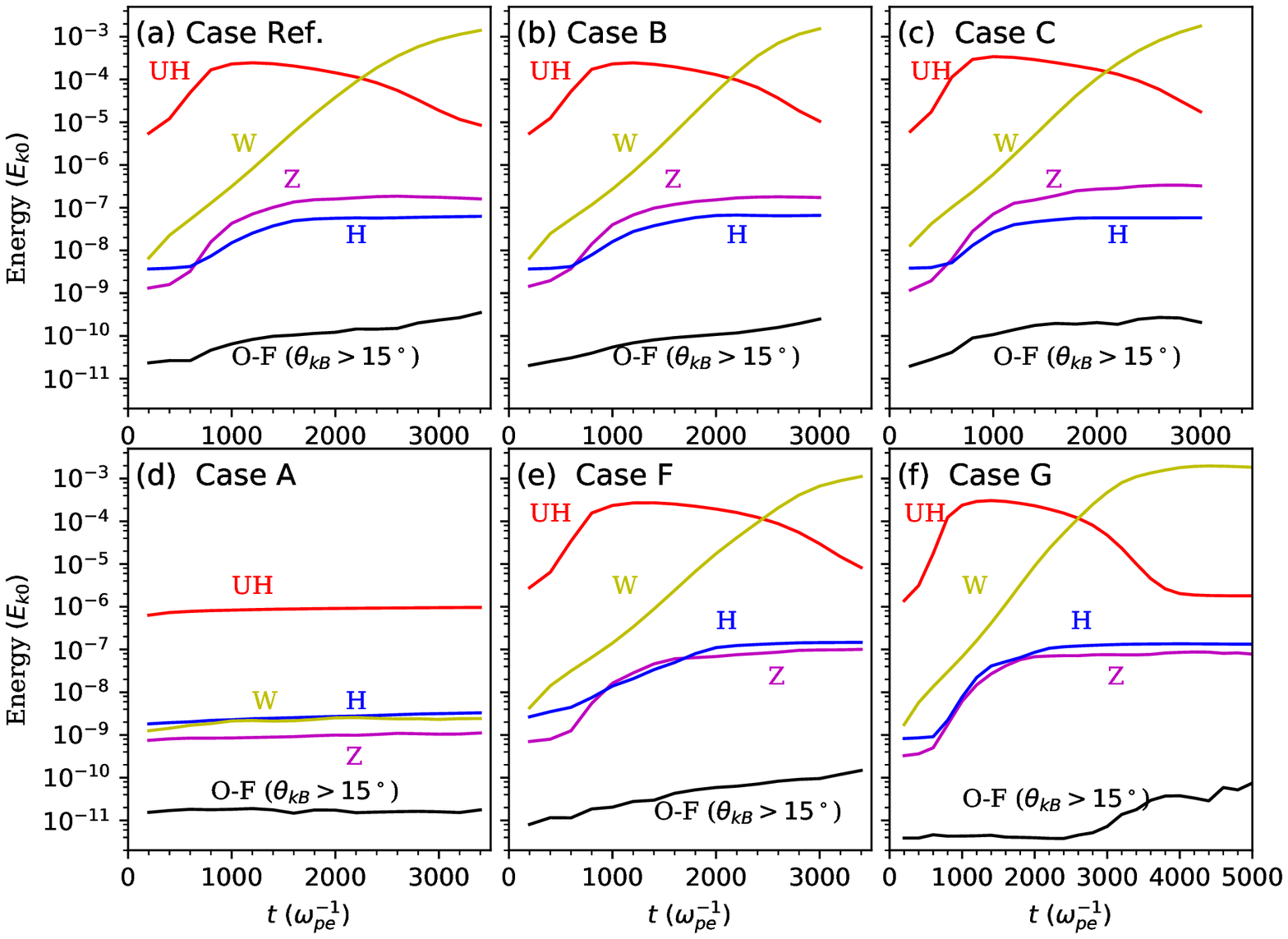}
              }
              \caption{The temporal energy profiles of UH, Z, W, H, O-F modes of different cases listed in Table \ref{Tab:Table1}.
              }
   \label{Fig:figure8}
   \end{figure*}

Comparisons of these cases with the reference case are presented in Figures \ref{Fig:figure7} and \ref{Fig:figure8}, where energy profiles of electro-magnetic field components and those of various wave modes are plotted. We reach the following conclusions:

\begin{enumerate}
  \item Resolving capability of the simulations:

  According to the setup of the reference case, the frequency and wave number ranges that can be resolved are [0, 32] $\Omega_{ce}$ and [-536, 536] $\Omega_{ce}/c$, respectively. The corresponding resolutions of the DFT analysis are 0.06 $\Omega_{ce}$ for the frequency ($\omega$) (for the time interval of 1000 $\omega_{pe}^{-1}$) and 1.04 $\Omega_{ce}/c$ for the wave number ($k$). The setup seems to be sufficient to properly resolve most modes except the O-F mode which is characterized by a range of relatively small wave number [0, 3] $\Omega_{ce}/c$. Thus, in the convergence test we paid special attention to the significance and variation of the O-F mode.

  \item Magnitudes and variation profiles of intensities:

  Intensities of all field components and relevant wave modes reach levels that are significantly higher than those in the Maxwellian case (A), even the O-F mode, the weakest one, is about one-order of magnitude higher than the level of the corresponding thermal noise.

  \item Effect of NPPC and other factors:

  As long as the NPPC is greater than 500, intensities of field components and wave modes (except the O-F mode) do not vary considerably with the domain size and cell size. With increasing NPPC, it takes more time for the O-F mode to rise towards the asymptotic intensity. To show its rising trend in Case G with NPPC = 4000, we extend the simulation time to 5000 $\omega_{pe}^{-1}$. Note that the O-F mode reaches similar intensities at the end of each simulation.

\end{enumerate}

Furthermore, to verify the validity of the underlying emission mechanism, especially, the generation of the O-F mode emission from the coalescence of Z and W modes, we conducted another study with the wave-pumping technique. The result is also obtained with the VPIC code. The dispersion analyses for three cases are examined, including the case (Z) in which only the Z mode has been pumped into the system, the case (W) in which only the W mode has been pumped, and the case (Z+W) in which both modes have been pumped. The results confirm robustly the generation of the O-F mode emission through the proposed ECMI-induced plasma emission process. The wave-pumping study will be published elsewhere and not shown here (\citealp{Ni2021}, submitted to Physics of Plasmas).

The above convergence test and additional study indicate that for wave modes except the O-F mode the quantitative results presented in the manuscript are convincing, while for the O-F mode the quantitative intensity evaluations may suffer from large uncertainties due to the limited resolving capability of its wave number, as well as the contamination from the spectrally-nearby Z mode. Thus, in the study we did not present the variation profiles of the O-mode intensity with $\omega_{pe}/\Omega_{ce}$.


\begin{table}[htbp]
\centering
\caption{Setup parameters of various cases for the convergence test. NPPC is the number of macro-particles per species per cell, $n_x$ and $n_z$ are numbers of grid cells, $dx$ and $dz$ are the cell size, and $\lambda_D$ is the Debye length of the background electrons. The reference case is taken to be the case presented in \S\ref{sec:3.2} with $\omega_{pe}/\Omega_{ce}=10$, $v_e=0.15c$, and NPPC = 1000. Parameters not listed are set to be the same as those of the reference case.}
\setlength{\tabcolsep}{5mm}{
\begin{tabular}{cccccc}
\hline
\hline
Case  & NPPC & $n_x$ & $n_z$ & $dx$ ($\lambda_D$) & $dz$ ($\lambda_D$) \\
\hline
Ref. & 1000 & 1024 & 1024 & 3.25 & 3.25 \\
A    & 1000 & 1024 & 1024 & 3.25 & 3.25 \\
B    & 1000 & 2048 & 2048 & 3.25 & 3.25 \\
C    & 250  & 2048 & 2048 & 3.25/2 & 3.25/2 \\
D    & 200  & 1024 & 1024 & 3.25 & 3.25 \\
E    & 500  & 1024 & 1024 & 3.25 & 3.25 \\
F    & 2000 & 1024 & 1024 & 3.25 & 3.25 \\
G    & 4000 & 1024 & 1024 & 3.25 & 3.25 \\
\hline
\end{tabular}
}
\label{Tab:Table1}
\end{table}



\begin{thebibliography}{}
\bibitem[Benz(2002)]{Benz2002} Benz, A.\ 2002, Plasma Astrophysics. Kinetic Processes in Solar and Stellar Coronae, second edition. By A. Benz, Institute of Astronomy, ETH Z{\"u}rich, Switzerland.  Astrophysics and Space Science Library, Vol. 279, Kluwer Academic Publishers, Dordrecht, 2002.. doi:10.1007/978-0-306-47719-5
\bibitem[Ben{\'a}{\v{c}}ek et al.(2017)]{Benacek17} Ben{\'a}{\v{c}}ek, J., Karlick{\'y}, M., \& Yasnov, L.~V.\ 2017, \aap, 598, A106. doi:10.1051/0004-6361/201629717
\bibitem[Ben{\'a}{\v{c}}ek \& Karlick{\'y}(2018)]{Benacek18} Ben{\'a}{\v{c}}ek, J., \& Karlick{\'y}, M.\ 2018, \aap, 611, A60. doi:10.1051/0004-6361/201731424
\bibitem[Ben{\'a}{\v{c}}ek \& Karlick{\'y}(2019)]{Benacek2019} Ben{\'a}{\v{c}}ek, J. \& Karlick{\'y}, M.\ 2019, \apj, 881, 21. doi:10.3847/1538-4357/ab2bfc

\bibitem[Bowers et al.(2009)]{Bowers2009} Bowers, K.~J., Albright, B.~J., Yin, L., et al.\ 2009, Journal of Physics Conference Series, 012055. doi:10.1088/1742-6596/180/1/012055
\bibitem[Bowers et al.(2008a)]{Bowers2008a} Bowers, K.~J., Albright, B.~J., Yin, L., et al.\ 2008a, Physics of Plasmas, 15, 055703. doi:10.1063/1.2840133
\bibitem[Bowers et al.(2008b)]{Bowers2008b} Bowers, K. J., Albright, B. J., Bergen, B., et al. 2008b, in SC '08: Proceedings of the 2008 ACM/IEEE conference on Supercomputing (New York: IEEE Press), 63, \url{http://dl.acm.org/citation.cfm?id=1413435}

\bibitem[Chen et al.(2011)]{Chen2011} Chen, B., T.~S. Bastian, D.~E. Gary, and J. Jing\ 2011,\ \apj, 736, 64. doi:10.1088/0004-637X/736/1/64
\bibitem[Chernov(1990)]{Chernov1990} Chernov, G.~P.\ 1990, \solphys, 130, 75. doi:10.1007/BF00156780
\bibitem[Chernov(2006)]{Chernov2006} Chernov, G.~P.\ 2006, \ssr, 127, 195. doi:10.1007/s11214-006-9141-7
\bibitem[Chernov(2011)]{Chernov2011} Chernov, G.\ 2011, Fine Structure of Solar Radio Bursts,  by Chernov, Gennady. ISBN: 978-3-642-20014-4. Berlin: Springer, 2011

\bibitem[Chernov et al.(2012)]{Chernov12} Chernov, G.~P., Sych, R.~A., Meshalkina, N.~S., et al.\ 2012, \aap, 538, A53. doi:10.1051/0004-6361/201117034
\bibitem[Chernov(2015)]{Chernov2015} Chernov, G.\ 2015, arXiv e-prints, arXiv:1512.06311

\bibitem[Dory et al.(1965)]{Dory65} Dory, R.~A., Guest, G.~E., \& Harris, E.~G.\ 1965, \emph{Physical Review Letters}, 14, 131. doi:10.1103/PhysRevLett.14.131
\bibitem[Ginzburg \& Zhelezniakov(1958)]{Ginzburg1958} Ginzburg, V.~L. \& Zhelezniakov, V.~V.\ 1958, \sovast, 2, 653
\bibitem[Guo et al.(2014)]{Guo2014} Guo, F., Li, H., Daughton, W., et al.\ 2014, \prl, 113, 155005. doi:10.1103/PhysRevLett.113.155005
\bibitem[Guo et al.(2015)]{Guo2015} Guo, F., Liu, Y.-H., Daughton, W., et al.\ 2015, \apj, 806, 167. doi:10.1088/0004-637X/806/2/167
\bibitem[Kaneda et al.(2017)]{Kaneda2017} Kaneda, K., Misawa, H., Iwai, K., et al.\ 2017, \apj, 842, 45. doi:10.3847/1538-4357/aa74c1
\bibitem[Karlick{\'y}(2013)]{Karlicky2013} Karlick{\'y}, M.\ 2013, \aap, 552, A90. doi:10.1051/0004-6361/201321356
\bibitem[Kasaba et al.(2001)]{Kasaba2001} Kasaba, Y., Matsumoto, H., \& Omura, Y.\ 2001, \jgr, 106, 18693. doi:10.1029/2000JA000329
\bibitem[Kuijpers(1975a)]{Kuijpers1975a} Kuijpers, J.\ 1975a, \solphys, 44, 173. doi:10.1007/BF00156854
\bibitem[Kuijpers(1975b)]{Kuijpers1975b} Kuijpers, J.\ 1975b, \aap, 40, 405
\bibitem[Kuznetsov(2005)]{Kuznetsov2005} Kuznetsov, A.~A.\ 2005, \aap, 438, 341. doi:10.1051/0004-6361:20052712
\bibitem[LaBelle et al.(2003)]{LaBelle2003} LaBelle, J., Treumann, R.~A., Yoon, P.~H., et al.\ 2003, \apj, 593, 1195. doi:10.1086/376732
\bibitem[Ledenev et al.(2006)]{Ledenev2006} Ledenev, V.~G., Yan, Y., \& Fu, Q.\ 2006, \solphys, 233, 129. doi:10.1007/s11207-006-2099-5
\bibitem[Li et al.(2019)]{Li2019} Li, C., Chen, Y., Kong, X., et al.\ 2019, \apj, 880, 31. doi:10.3847/1538-4357/ab270f
\bibitem[Ni et al.(2020)]{Ni2020} Ni, S., Chen, Y., Li, C., et al.\ 2020, \apjl, 891, L25. doi:10.3847/2041-8213/ab7750
\bibitem[Ni et al.(2021)]{Ni2021} Ni, S., Chen, Y., Li, C., et al.\ 2021, An alternative form of the fundamental plasma emission through the coalescence of Z-mode waves with whistlers, Physics of Plasmas, submitted.

\bibitem[Slottje(1972)]{Slottje1972} Slottje, C.\ 1972, \solphys, 25, 210. doi:10.1007/BF00155758
\bibitem[Smerd \& Dulk(1971)]{Smerd1971} Smerd, S.~F. \& Dulk, G.~A.\ 1971, Solar Magnetic Fields, 43, 616
\bibitem[Stewart(1985)]{Stewart1985} Stewart, R.~T.\ 1985, Solar Radiophysics: Studies of Emission from the Sun at Metre Wavelengths, 361
\bibitem[Tan(2010)]{Tan2010} Tan, B.\ 2010, \apss, 325, 251. doi:10.1007/s10509-009-0193-5
\bibitem[Tan et al.(2014a)]{Tan14} Tan, B., Tan, C., Zhang, Y., et al.\ 2014a, \apj, 780, 129. doi:10.1088/0004-637X/780/2/129
\bibitem[Tan et al.(2014b)]{Tan2014} Tan, B., Tan, C., Zhang, Y., et al.\ 2014b, \apj, 790, 151. doi:10.1088/0004-637X/790/2/151
\bibitem[Thurgood \& Tsiklauri(2015)]{Thurgood2015} Thurgood, J.~O. \& Tsiklauri, D.\ 2015, \aap, 584, A83. doi:10.1051/0004-6361/201527079
\bibitem[Umeda(2010)]{Umeda2010} Umeda, T.\ 2010, Journal of Geophysical Research (Space Physics), 115, A01204. doi:10.1029/2009JA014643
\bibitem[Vasanth et al.(2019)]{Vasanth2019} Vasanth, V., Chen, Y., Lv, M., et al.\ 2019, \apj, 870, 30. doi:10.3847/1538-4357/aaeffd
\bibitem[Vasanth et al.(2016)]{Vasanth2016} Vasanth, V., Chen, Y., Feng, S., et al.\ 2016, \apjl, 830, L2. doi:10.3847/2041-8205/830/1/L2
\bibitem[Villasenor \& Buneman(1992)]{Villasenor1992} Villasenor, J. \& Buneman, O.\ 1992, Computer Physics Communications, 69, 306. doi:10.1016/0010-4655(92)90169-Y

\bibitem[Vlahos et al.(1982)]{Vlahos1982} Vlahos, L., Gergely, T.~E., \& Papadopoulos, K.\ 1982, \apj, 258, 812. doi:10.1086/160128

\bibitem[Wild \& Smerd(1972)]{Wild1972} Wild, J.~P. \& Smerd, S.~F.\ 1972, \araa, 10, 159. doi:10.1146/annurev.aa.10.090172.001111
\bibitem[Winglee \& Dulk(1986)]{Winglee86} Winglee, R.~M., \& Dulk, G.~A.\ 1986, \apj, 307, 808. doi:10.1086/164467
\bibitem[Wu \& Lee(1979)]{Wu1979} Wu, C.~S., \& Lee, L.~C.\ 1979, \apj, 230, 621. doi: 10.1086/157120
\bibitem[Yasnov \& Karlick{\'y}(2004)]{Yasnov04} Yasnov, L.~V., \& Karlick{\'y}, M.\ 2004, \solphys, 219, 289. doi:10.1023/B:SOLA.0000022942.17621.88
\bibitem[Yin et al.(2007)]{Yin2007} Yin, L., Albright, B.~J., Bowers, K.~J., et al.\ 2007, \prl, 99, 265004. doi:10.1103/PhysRevLett.99.265004

\bibitem[Zheleznyakov et al.(2016)]{Zheleznyakov2016} Zheleznyakov, V.~V., Zlotnik, E.~Y., Zaitsev, V.~V., et al.\ 2016, Physics Uspekhi, 59, 997. doi:10.3367/UFNe.2016.05.037813
\bibitem[Zheleznyakov \& Zlotnik(1975)]{Zheleznyakov1975} Zheleznyakov, V.~V., \& Zlotnik, E.~Y.\ 1975, \solphys, 44, 461. doi:10.1007/BF00153225
\bibitem[Zhou et al.(2020)]{Zhou2020} Zhou, X., Mu{\~n}oz, P.~A., B{\"u}chner, J., et al.\ 2020, \apj, 891, 92. doi:10.3847/1538-4357/ab6a0d
\bibitem[Ziebell et al.(2015)]{Ziebell2015} Ziebell, L.~F., Yoon, P.~H., Petruzzellis, L.~T., et al.\ 2015, \apj, 806, 237. doi:10.1088/0004-637X/806/2/237
\bibitem[Zlotnik et al.(2014)]{Zlotnik2014} Zlotnik, E.~Y., Zaitsev, V.~V., \& Altyntsev, A.~T.\ 2014, \solphys, 289, 233. doi:10.1007/s11207-013-0327-3
\bibitem[Zlotnik(2013)]{Zlotnik13} Zlotnik, E.~Y.\ 2013, \solphys, 284, 579. doi:10.1007/s11207-012-0151-1

\end{thebibliography}
\end{document}